\documentclass[amsmath,amssymb,10pt,reprint,notitlepage,showpacs,superscriptaddress,longbibliography]{revtex4-1}
\usepackage{graphicx}
\usepackage{calc}
\usepackage{braket}
\usepackage{ulem}   
\usepackage{slashed}
\usepackage{wasysym}
\usepackage{siunitx}
\usepackage{dsfont}
\usepackage{amsthm,amsmath,amsfonts,amssymb,verbatim,color}
\usepackage{graphicx}
\usepackage{subfigure}
\usepackage{bm}
\usepackage{epsfig,slashed}
\usepackage[T1]{fontenc}
\usepackage{ulem}
\usepackage[colorlinks=true,citecolor=blue,linkcolor=blue,urlcolor=blue]{hyperref}
\usepackage[version=4]{mhchem}
\usepackage{multirow}
\usepackage{soul}
\usepackage{caption}
\begin{document}
\title{Magneto-optical effects of an artificially-layered ferromagnetic topological insulator with T$_C$ of 160 K}

\author{Xingyue Han}
\affiliation{Department of Physics and Astronomy, University of Pennsylvania, Philadelphia, Pennsylvania 19104, USA}
\author{Hee Taek Yi}
\affiliation{Department of Physics and Astronomy, Rutgers, The State University of New Jersey, Piscataway, New Jersey 08854, USA}
\author{Seongshik Oh}
\affiliation{Department of Physics and Astronomy, Rutgers, The State University of New Jersey, Piscataway, New Jersey 08854, USA}
\author{Liang Wu}
\email{liangwu@sas.upenn.edu}
\affiliation{Department of Physics and Astronomy, University of Pennsylvania, Philadelphia, Pennsylvania 19104, USA}

\begin{abstract}
\textbf{Abstract:}
Magnetic topological insulator is a fertile platform to study the interplay between magnetism and topology. The unique electronic band structure can induce exotic transport and optical properties. However, a comprehensive optical study in both near-infrared frequency and terahertz frequency has been lacking. Here, we report magneto-optical effects from a heterostructure of Cr-incorporated topological insulator, CBST. We use 800 nm magneto-optical Kerr effect to reveal a ferromagnetic order in the CBST film with a high transition temperature at 160 K. We also use time-domain terahertz polarimetry to reveal a terahertz Faraday rotation of 1.5 mrad and Kerr rotation of 5.1 mrad at 2 K. The calculated terahertz Hall conductance is 0.42 $e^2/h$.  Our work shows the optical responses of an artificially layered  magnetic topological insulator, paving the way towards high-temperature quantum anomalous Hall effect via heterostructure engineering.

\textbf{key words:}  magnetic topological insulators, magneto-optical Kerr effect, terahertz Faraday rotation, terahertz Kerr rotation
\end{abstract}

\maketitle

Topological insulators (TIs), characterised by the insulating bulk states and conducting topological surface states, were originally proposed in time-reversal-invariant systems \cite{moore2010Nat, hasan2010RMP, qi2011RMP, tokura2019Nat}. The gapless topological surface states are located in the bulk gap crossing at the Dirac point. Incorporating ferromagnetism into TIs breaks time-reversal symmetry and opens an exchange gap at the Dirac point \cite{yu2010Sci, chang2013Sci, tokura2019Nat}. The interplay between magnetism and topology has brought a plethora of exotic phenomena \cite{onoda2003PRL, wan2011PRB, nakatsuji2015Nat, armitage2018RMP, tokura2019Nat, yu2010Sci}, including the quantum anomalous Hall effect (QAHE) \cite{chang2013Sci}, and the axion insulator states \cite{liu2020NatMat}.  When the Fermi energy lies in the exchange gap, the QAHE emerges due to dissipationless chiral edge modes. In an ideal quantum anomalous Hall state, the anomalous Hall resistance is quantized at $h/e^2$, while the longitudinal resistivity vanishes. In order to realize a ferromagnetic TI system, the first attempt was doping nonmagnetic TIs with magnetic impurities, such as Cr- and V- doped (Bi,Sb)$_2$Te$_3$ \cite{chang2013Sci, chang2015NatMat, checkelsky2014NatPhy}. However, the inhomogeneity of the magnetic dopants suppresses the effective energy gap. It results in a low observation temperature of QAHE, usually below 100 mK. Following the magnetic impurity doping method, the magnetic modulation doping technique increased the surface mass gap and improved the homogeneity. The QAHE observation temperature was raised to 2 K \cite{mogi2015APL, mogi2017NatMat, mogi2017SciAdv, ou2018AdvMat, Yi2023NanoLetter}. Recently, intrinsic magnetic TIs with chemical formulas such as MnBi$_2$Se$4$ \cite{otrokov2019nature, deng2020Sci, Li2019SciAdv} have received more attention with the report of 97\% quantization persist up to 6.5 K in a 5 septuple-layer (SL) flake\cite{deng2020Sci}.

The tunability of the magnetic and topological properties of magnetic TIs have been investigated in various ways, such as Sb doping \cite{Guan2022PRM, hu2021PRB}, hydrostatic pressure \cite{eckberg2023arxiv, Chen2019PRM}, and surface engineering \cite{mazza2022AFM}. Another important method is using nonmagnetic spacer layer \cite{yao2021Nano}. For example, in the intrinsic \ce{MnBi2Te4} family, due to the van der Waals nature, the magnitude of interlayer antiferromagnetic strength can be suppressed by inserting \ce{Bi2Te3} quintuple layers (QL) between \ce{MnBi2Te4} SLs, with the chemical formula MnBi$_{2n}$Te$_{3n+1}$. The electric band structure spontaneously changes with different magnetic orders. For example, \ce{MnBi2Te4} (n=1) \cite{deng2020Sci} and \ce{MnBi4Te7} (n=2) \cite{hu2021PRB} are reported to be antiferromagnetic TIs, \ce{MnBi8Te13} (n=4) is predicted to be ferromagnetic TI \cite{hu2020SciAdv, zhong2021nanoletter}. With the boost of research interest in the MnBi$_{2n}$Te$_{3n+1}$ family, replacing Mn with other magnetic elements has been investigated as well. Density functional theory (DFT) calculations revealed that certain elements can form stable \ce{XBi2Te4} SL structures, including X = Ti, V, Ni, Eu, etc \cite{Li2019SciAdv}. But X = Cr, Fe, and Co show instability to form SL structures \cite{Li2019SciAdv}.

\begin{figure}
\includegraphics[width=0.4\textwidth]{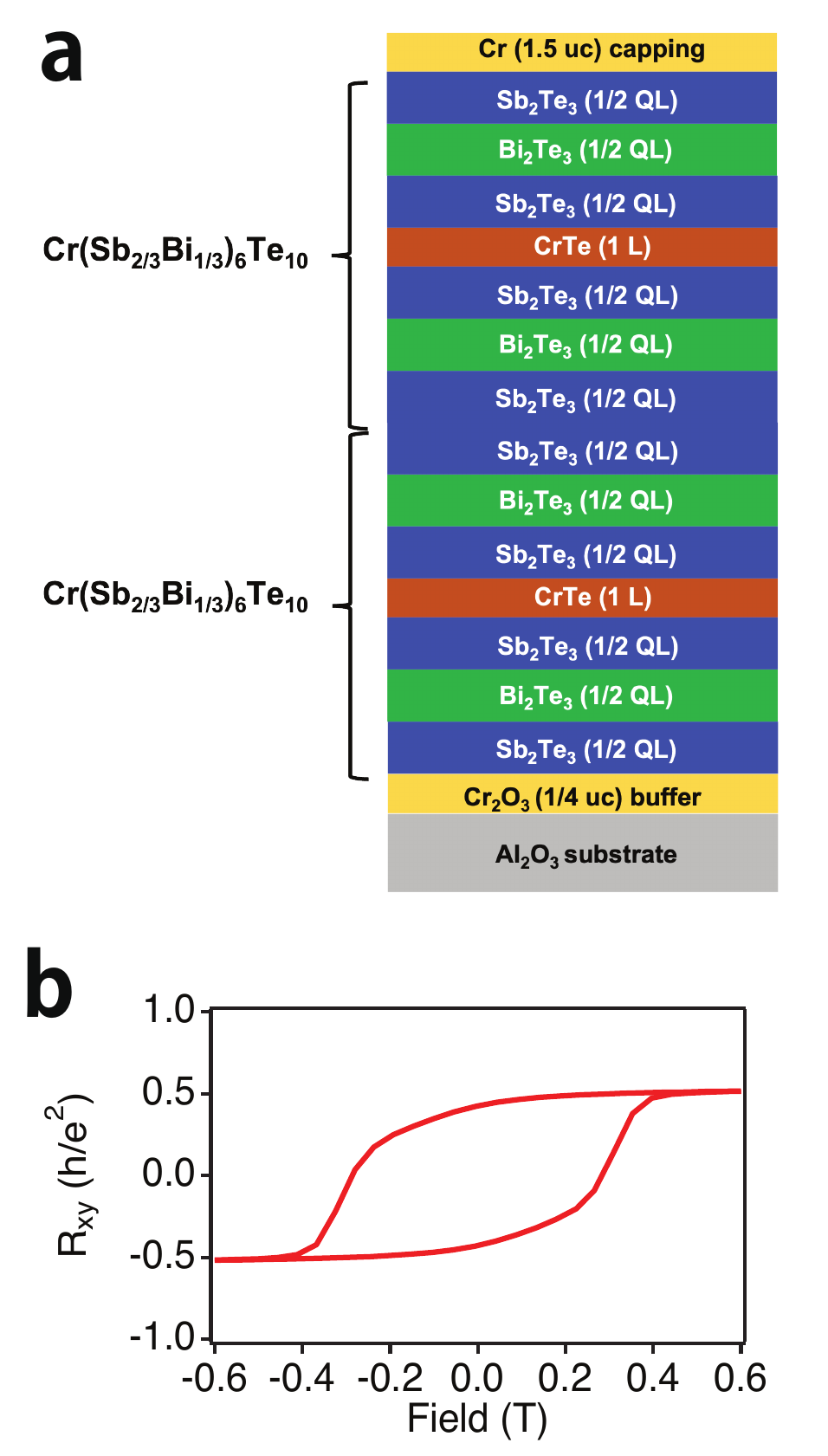}
\caption{\textbf{Structure and transport of CBST.} \textbf{a.} Layered structure of CBST. The 17-tuple layer structure is repeated twice. The film terminates at BST layer. \textbf{b.} Magnetic field dependence of Hall resistance of CBST measured at 2 K.  }
\label{Fig1}
\end{figure}

Heterostructure engineering is a powerful tool to tune the magnetic and topological properties of Cr-incorporated TIs \cite{jiang2020NatMat,mogi2022NatPhy,yasuda2016NatPhy, yao2021Nano}. In our work, we report a heterostructure of Cr-incorporated topological insulator, dubbed as CBST. The layer-by-layer epitaxial deposition of each \ce{Sb2Te3}, \ce{Bi2Te3} and CrTe was carried out based on the Cr(Bi$_{1/3}$Sb$_{2/3}$)$_6$Te$_{10}$ structure (Fig.\ref{Fig1}\textbf{a}).  We performed magneto-optical measurements in both near-infrared (NIR) and terahertz (THz) frequencies. Ferromagnetic order is observed using magneto-optical Kerr effect (MOKE). The magnetic transition temperature at 160 K is higher than any other Cr-incorporated magnetic TIs. We also performed THz measurements in CBST which exhibits a Faraday rotation of 1.5 mrad and Kerr rotation of 5.1 mrad under 6 T at 2 K.

In the CBST film, the Cr(Bi$_{1/3}$Sb$_{2/3}$)$_6$Te$_{10}$ structure repeats twice, terminating at (Bi$_{1/2}$Sb$_{1/2}$)$_2$Te$_3$ (BST) layer (Fig.\ref{Fig1}\textbf{a}). The 6.5 nm-thick film is grown on a 1 cm $\times$ 1 cm \ce{Al2O3} (0001) substrate using a custom-built molecular beam epitaxy (MBE) system (SVTA) under a base pressure of 5 $\times$ 10$^{-10}$ Torr. High-purity (99.999\%) Bi,Sb, Te, and Cr were thermally evaporated using effusion cells, and all the source fluxes were calibrated in-situ by quartz crystal micro-balance and ex-situ by Rutherford backscattering spectroscopy. We employ thin epitaxial \ce{Cr2O3} as a buffer layer to assist the topological layers in adhering to the \ce{Al2O3} substrate, and naturally oxidized amorphous CrO$_x$ as a capping layer to suppress degradation and boost ferromagnetic order \cite{yao2021Nano, Yi2023NanoLetter}. High-angle annular dark-field scanning transmission electron microscope (HAADF-STEM) image has revealed that Cr is incorporated into the neighboring Sb$_2$Te$_3$ layer instead of forming a van der Waals gap between Sb$_2$Te$_3$ layers \cite{yao2021Nano}. The transport data at 2 K (Fig.\ref{Fig1}\textbf{b}) with single coercive field inicates that surface states only form at top and bottom surfaces, suggesting that the entire system is behaving as a single magnetic TI layer.

\begin{figure}
\includegraphics[width=0.5\textwidth]{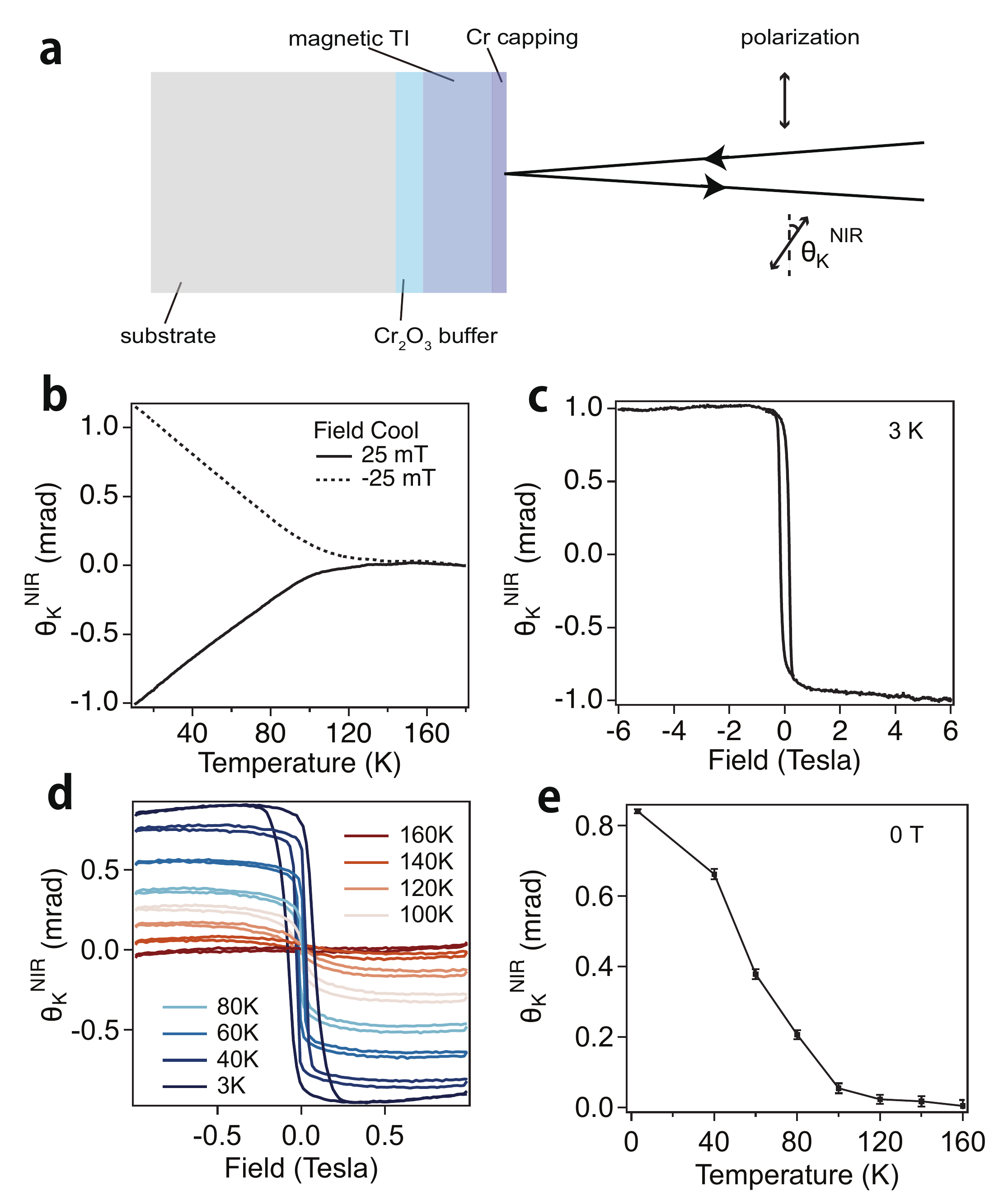}
\caption{\textbf{MOKE results of CBST at 800 nm.} \textbf{a.} Schematic of MOKE measurement from CBST. \textbf{b.} Temperature dependent MOKE under field-cool. $\pm$25 mT magnetic field is applied perpendicular to the sample surface. \textbf{c.} MOKE Hysteresis loop measured at 3 K. \textbf{d.} MOKE hysteresis loops measured from 3 K to 160 K. \textbf{e.} Temperature dependence of the field-symetrized $\theta_K$ at 0 T.}
\label{Fig2}
\end{figure}

We firstly utilize MOKE to probe the time-reversal symmetry breaking in CBST. A mode-locked Ti:sapphire laser with 800 nm centre wavelength, 80 MHz repetition rate and 50 fs pulse duration is used to measure the Kerr rotation $\theta_K^{NIR}$ from the CBST sample \cite{xu2022NatPhy}. $\theta_K^{NIR}$ is the polarization plane change in the reflection beam proportional to the net magnetization along the laser propagation direction \cite{mccord2015JPD} (Fig.\ref{Fig2}\textbf{a}). We carried out temperature dependent MOKE measurements under an out-of-plane magnetic field at $\pm$ 25 mT. As shown in Fig.\ref{Fig2}\textbf{b}, cooling from 180 K to 10 K, $\theta_K^{NIR}$ emerges around 160 K and shows opposite signs under opposite magnetic fields.  Fig.\ref{Fig2}\textbf{c} and \textbf{d} show the magnetic field dependent MOKE at 3 K and higher temperatures. $\theta_K^{NIR}$ at 3 K shows a hysteresis loop with the coercive field H$_C$ at 70 mT and saturates at 1 mrad. Above the saturation field, there is no signs of spin-flip/flop transitions up to $\pm$ 6 T, consistent with the behavior from a ferromagnetic order. The coercive field and MOKE signal monotonically decrease with increasing temperature, and disappears at 160 K (Fig.\ref{Fig2}\textbf{d}). Fig.\ref{Fig2}\textbf{e} shows zero-field MOKE 
\begin{equation}
\theta^{NIR}_K (0T) = \frac{\theta^{NIR}_K (+0T) - \theta^{NIR}_K(-0T)}{2}
\end{equation} 
under different temperatures.  The high transition temperature indicates an enhanced ferromagnetic order in the CBST film, compared to other Cr-incorporated magnetic TIs (TABLE.\ref{table1}), achieved by the high Cr concentration from the layer-by-layer growth method and an active capping layer CrO$_x$ \cite{yao2021Nano, Yi2023NanoLetter}. Note that our table does not include other dopants to compare different growth method and heterostructures \cite{li2016scirep, chang2015NatMat, Liu2017PRL} or heavily ferromagnetic doped bulk samples due to possible clustering \cite{Wimmer2021AdvMat, tokura2019Nat} .

\begin{table}[h!]
\centering
\begin{tabular}{||c|c c||} 
 \hline
\multirow{1}{5em}{Type} & Sample & T$_N$/T$_C$ \\ [0.5ex] 
 \hline\hline
 \multirow{3}{5em}{Intrinsic} & \ce{MnBi2Te4}\cite{gong2019CPL, zhao2021Nano, liu2021Arx, LIU2022JCG} & 15-25 K\\ 
 &\ce{MnBi6Te10}\cite{yan2022NanoLet} & 13 K\\
 &\ce{MnBi8Te13}\cite{hu2020SciAdv} & 10 K\\
 \hline
 \multirow{3}{5em}{Uniform doping} & Cr$_{0.15}$(Bi$_{0.1}$Sb$_{0.9}$)$_{1.85}$Te$_3$ \cite{chang2013Sci} & 15 K\\
 &Cr$_{0.22}$(Bi$_{0.2}$Sb$_{0.8}$)$_{1.78}$Te$_3$ \cite{checkelsky2014NatPhy} & 45 K\\
 &Cr$_{0.29}$Sb$_{1.71}$Te$_3$ \cite{chang2015NatMat} & 59 K\\
 \hline
 \multirow{2}{5em}{Modulation doping} & Cr$_{0.57}$(Bi$_{0.26}$Sb$_{0.74}$)$_{1.43}$Te$_3$ \cite{okada2016NatComm} & 70 K\\
 & Cr$_{0.95}$(Bi$_{0.32}$Sb$_{0.68}$)$_{1.05}$Te$_3$ \cite{mogi2015APL} & 80 K\\
 \hline
\multirow{1}{5em} & layered heterostructure (this work) & 160 K\\
 \hline
\end{tabular}
\caption{Magnetic transition temperatures of MnBi$_{2n}$Te$_{3n+1}$ family and Cr-incorporated magnetic TIs. }
\label{table1}
\end{table}

\begin{figure}
\includegraphics[width=0.5\textwidth]{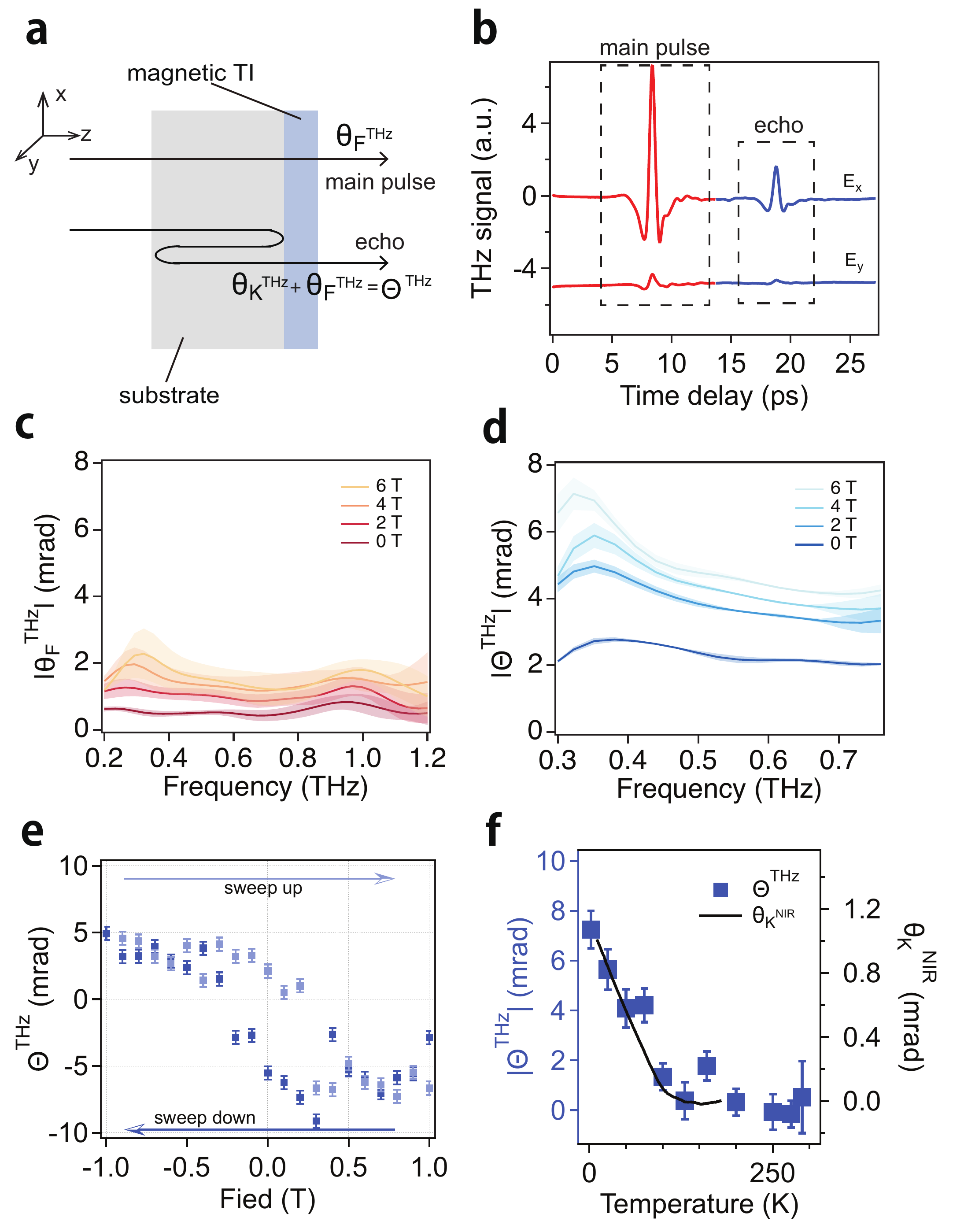}
\caption{ \textbf{THz result of CBST.} \textbf{a.} Schematics of the polarization state of the transmitted THz waves after CBST. The incident THz polarization is parallel to x-axis.  \textbf{b.} Terahertz signals in time domain for the CBST. The transmitted light is composed of the main pulse and the first echo. They are separated in time domain due to different light path. E$_x$ and E$_y$ are the transmitted signals parallel and perpendicular, respectively, to the incident polarization at 0 T. \textbf{c.} Spectra of THz Faraday rotation at 2 K. \textbf{d.} Spectra of $\Theta^{THz}$, the sum of THz Faraday rotation $\theta_{\mathrm{F}}^{\mathrm{THz}}$ and THz Kerr rotation, at 2 K. The shaded area in \textbf{c} and \textbf{d} are error bars. \textbf{e.} Field dependence of $\Theta^{THz}$ at 2 K. Each data point is calculated from the average value in the range 0.2 THz to 1.4 THz. \textbf{f.} Temperature dependence of $\Theta^{THz}$ (blue squares) and $\theta_K^{NIR}$ (black curve). }
\label{Fig3}
\end{figure}

Then we study the low-energy response of CBST film  by transmission time-domain terahertz (THz) polarimetry from 0.2 THz (0.8 meV) to 1.2 THz (5.0 meV) \cite{Han2022PRB}. A pair of photoconductive antennas are used to generate and probe THz radiations. A delay stage introducing time delay between THz emission and the detection path enables the time-domain measurement. A THz polarizer mounted on a rotation stage is used to analyze THz polarization transmitted through the CBST film. In magnetic TIs, the exchange energy gap opening at the TI surface is usually tens of meV \cite{lee2015PNAS,chen2010Sci}. Thus interband transition is prohibited under THz excitation. We then measure the THz Faraday $\theta_{\mathrm{F}}^{\mathrm{THz}}$ and Kerr rotation $\theta_{\mathrm{K}}^{\mathrm{THz}}$ from CBST film. By definition, Faraday rotation refers to the polarization change of the transmitted light, while Kerr rotation is about the reflected light. In the time-domain spectroscopy, the transmitted light is composed of the main pulse and a series of echos which are separated in time (Fig.\ref{Fig3}\textbf{b}). The main pulse is the directly transmitted THz wave (Fig.\ref{Fig3}\textbf{a}). The first echo is firstly reflected at the substrate/TI interface, then reflected back at the substrate surface, finally transmitted through the film (Fig.\ref{Fig3}\textbf{a}). Higher orders of echos  appear later in time, and are not being considered here. Therefore, the polarization change of the main pulse is the Faraday rotation $\theta_{\mathrm{F}}^{\mathrm{THz}}$ (Fig.\ref{Fig3}\textbf{a,b}). The polarization change of the first echo is the sum of Faraday rotation and Kerr rotation $\Theta^{THz}=\theta_{\mathrm{F}}^{\mathrm{THz}}+\theta_{\mathrm{K}}^{\mathrm{THz}}$ (Fig.\ref{Fig3}\textbf{a,b}). Here we use the superscript "THz" to distinguish the THz rotations with the NIR MOKE results in Fig.\ref{Fig2}. By fixing the incident polarization (x direction), the THz rotation angles can be calculated from the ratio of the perpendicular component ($E_y$) and the parallel component ($E_x$).

To measure the field-dependence of $\theta_{\mathrm{F}}^{\mathrm{THz}}$ and $\Theta^{THz}$, the magnetic field is swept to 6 T to magnetize the film before measurements. A sequence of 6 T $\rightarrow$ 4 T $\rightarrow$ 2 T $\rightarrow$ 0 T $\rightarrow$ -6 T $\rightarrow$ -4 T $\rightarrow$ -2 T $\rightarrow$ 0 T is followed to apply the field. Fig.\ref{Fig3}\textbf{c} displays the field-symmetrized THz Faraday and Kerr rotation spectra
\begin{equation}
\theta_{\mathrm{F}}^{\mathrm{THz}}(B) = \frac{\theta_{\mathrm{F}}^{\mathrm{THz}}(+B)-\theta_{\mathrm{F}}^{\mathrm{THz}}(-B)}{2}
\end{equation} 
 at 2 K. A modest frequency dependence agrees with previous THz measurements in magnetic TIs \cite{okada2016NatComm, mogi2022NatPhy}. Under 0 T, $\theta_{\mathrm{F}}^{\mathrm{THz}}$ shows an average value of 0.57 mrad. It increases monotonically with magnetic field and reaches 1.5 mrad under 6 T. Field-symmetrized spectra of $\Theta^{THz}$ in the range 0.3 THz (1.2 meV) to 0.76 THz (3.1 meV) are displayed in Fig.\ref{Fig3}\textbf{d}. The more substantial frequency dependence of $\Theta^{THz}$ at high fields comes from the temporal overlap between the main pulse and the echo \cite{okada2016NatComm}. The average value of $\Theta^{THz}$ is 2.3 mrad under 0 T and 5.1 mrad under 6 T. The field and temperature dependence of $\Theta^{THz}$ are displayed in Fig.\ref{Fig3}\textbf{e} and \textbf{f}. A hysteresis loop is observed in $\Theta^{THz}$, and its temperature dependence well agrees with $\theta_K^{NIR}$. These observations corroborates the magnetic origin of the THz rotations in CBST.

\begin{figure}
\includegraphics[width=0.5\textwidth]{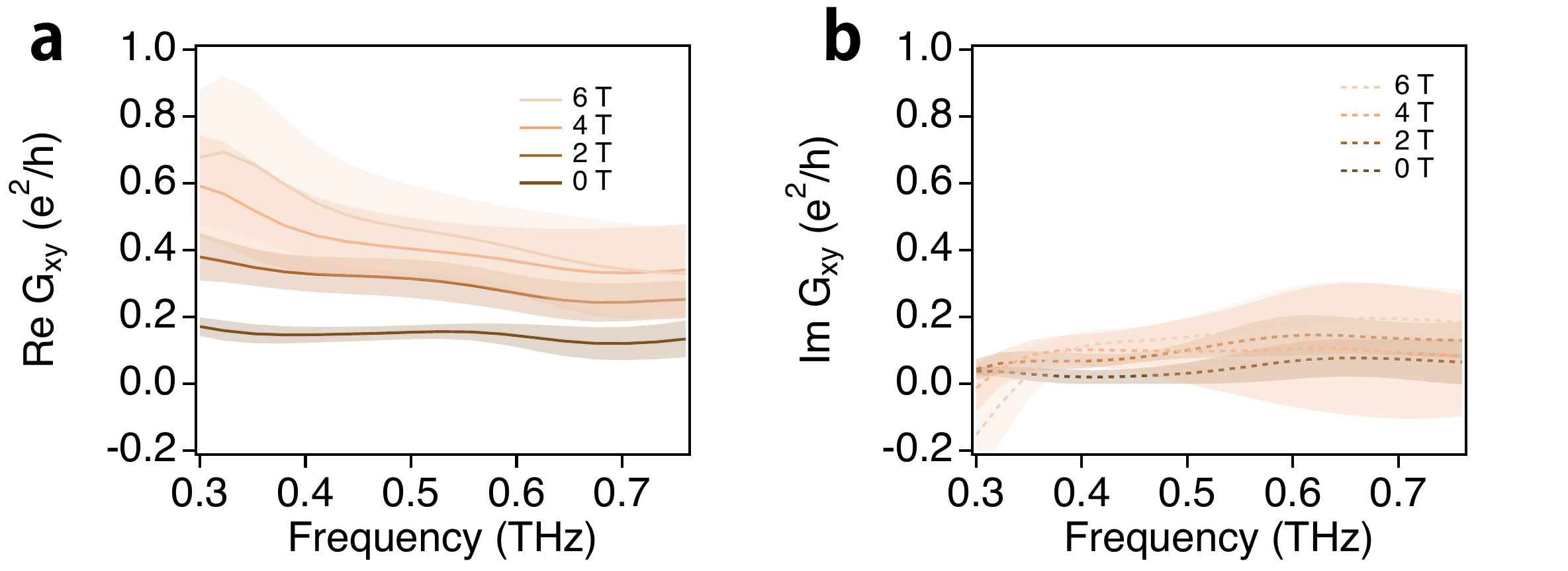}
\caption{\textbf{THz anomalous Hall conductance of CBST. } $G_{xy}$ is the complex anomalous Hall conductivity. \textbf{a.} and \textbf{b.} shows the real and imaginary part of the 2K $G_{xy}$, respectively. Shaded area is the error bar.}
\label{Fig4}
\end{figure}

THz Faraday rotation is related to THz Hall conductance $G_{xy}$ via

\begin{equation}
G_{xy}=\theta_{\mathrm{F}}^{\mathrm{THz}}(G_{xx}+\dfrac{n_{s}+1}{Z_0})
\end{equation} 

 Here $G_{xx}$ is the longitudinal conductance, $n_s$ is the refractive index of the substrate (3.1 for sapphire), and $Z_0$ ($\approx$376.7 $\Omega$) is the vacuum impedance. We obtain a small $G_{xx}$ in the probing frequency range from the nearly unity THz transmission $T_{xx}$, according to 

 \begin{equation}
G_{xx}=\dfrac{n_s+1}{Z_0}[\dfrac{e^{i\omega(n_s-1)\Delta L/c}}{T_{xx}}-1]
\end{equation}

Here $c$ is the speed of light, and $\Delta L$ is the thickness mismatch between the CBST sample and reference substrate. 
As shown in Fig.\ref{Fig4}\textbf{a.}, the Hall conductance is almost constant at 0.14 $e^2/h$ under 0 T. Under higher magnetic fields, $G_{xy}$ decreases at higher frequencies. Under 6 T, $G_{xy}$ reaches a maximum value of 0.69 $e^2/h$ at 0.32 THz, and converges to 0.33 $e^2/h$ at 0.7 THz. The average value of the 6 T spectrum is 0.42 $e^2/h$. The frequency dependence at high field might come from cyclotron resonances\cite{wu2016Sci, wu2015PRL}. A small but nonzero value in the imaginary part of the complex $G_{xy}$ (Fig.\ref{Fig4}\textbf{b.}) also indicates the emergence of dissipation.
 
 In the ideal QAHE state, $G_{xx}$ vanishes to zero, while $G_{xy}$ is quantized in unit of $e^2/h$. The calculated THz Faraday and Kerr rotations are $tan^{-1}(\frac{2\alpha}{n_s+1})=3.67$ mrad and $tan^{-1}(\frac{4n_s\alpha}{n_s^2-1})=10.5$ mrad, respectively \cite{wu2016Sci,okada2016NatComm, mogi2022NatPhy}. To account for the deviation between experimental results and theoretical calculations, different reasons were proposed such as trivial bands involvement \cite{li2016scirep, pan2020SciAdv} and emergence of superparamagnetic order \cite{lachman2015SciAdv, pan2020SciAdv}. Further investigation in CBST about the Fermi level and possible superparamagnetic order are needed towards the realization of QAHE . Looking forward, we hope this work generate interests in realizing high-temperature QAHE at terahertz frequency.\\

\textbf{Acknowledgments}
This project at Penn and Rutgers is mainly sponsored by the Army Research Office and was accomplished under Grant Number W911NF-20-2-0166.  X.H. is also partially supported by the Gordon and Betty Moore Foundation’s EPiQS Initiative, Grant GBMF9212 to L.W. and the NSF EPM program under grant no. DMR-2213891 for manuscript writing.  The work at Rutgers by H. Yi and S. O. was also supported by NSF DMR2004125, and the center for Quantum Materials Synthesis (cQMS), funded by the Gordon and Betty Moore Foundation’s EPiQS initiative through grant GBMF10104. It is supported in part by grant NSF PHY-1748958 to the Kavli Institute for Theoretical Physics (KITP).

\bibliographystyle{achemso}
\bibliography{CBST_V4}

\pagebreak


\end{document}